\RequirePackage[mathlines,pagewise]{lineno}
\documentclass[prd,twocolumn,amsmath,amssymb,showpacs,superscriptaddress,nofootinbib]{revtex4-1}
\usepackage[english]{babel}
\usepackage{graphicx,epsfig}
\usepackage{dcolumn}
\usepackage{bm,color}
\usepackage{multirow}
\usepackage[dvipdfm,CJKbookmarks=true,unicode,colorlinks,linkcolor=blue,anchorcolor=blue,citecolor=blue,pdfborder={0 0 0}]{hyperref}
\usepackage[pagewise]{lineno}

\graphicspath{{figures/}}

\usepackage{overpic,graphicx}
\usepackage{dcolumn}
\usepackage{bm}
\usepackage{rotating}
\usepackage{amsfonts}
\usepackage{keyval,graphicx}
\usepackage{textcomp,wasysym}
\usepackage{subfigure}
\usepackage{float}
\graphicspath{{figures/}}

\begin{document}
\hyphenation{BESII}
\hyphenation{BESIII}
\hyphenation{BEPCII}

\newcommand{\chisq}[1]{$\chi^{2}_{#1}$}
\newcommand{\GeV}{GeV/$c^2$}
\newcommand{\MeV}{MeV/$c^2$}
\newcommand{\br}[1]{\mathcal{B}(#1)}
\newcommand{\etap}{\eta^\prime}
\newcommand{\psip}{\psi^{\prime}}
\newcommand{\ar}{\rightarrow}

\title{\Large \boldmath \bf Observation of the Dalitz Decay $\eta' \to \gamma e^+e^-$}

\author{
  \begin{small}
    \begin{center}
      M.~Ablikim$^{1}$, M.~N.~Achasov$^{9,a}$, X.~C.~Ai$^{1}$,
      O.~Albayrak$^{5}$, M.~Albrecht$^{4}$, D.~J.~Ambrose$^{44}$,
      A.~Amoroso$^{48A,48C}$, F.~F.~An$^{1}$, Q.~An$^{45}$,
      J.~Z.~Bai$^{1}$
% ...
      \\
      \vspace{0.2cm}
      (BESIII Collaboration)\\
      \vspace{0.2cm} {\it
        $^{1}$ Institute of High Energy Physics, Beijing 100049, People's Republic of China\\
% ...
$^{4}$ Bochum Ruhr-University, D-44780 Bochum, Germany\\
        $^{5}$ Carnegie Mellon University, Pittsburgh, Pennsylvania 15213, USA\\
% ...
$^{9}$ G.I. Budker Institute of Nuclear Physics SB RAS (BINP), Novosibirsk 630090, Russia\\
% ...
$^{44}$ University of Rochester, Rochester, New York 14627, USA\\
        $^{45}$ University of Science and Technology of China, Hefei 230026, People's Republic of China\\
% ...
$^{48}$ (A)University of Turin, I-10125, Turin, Italy; (B)University of Eastern Piedmont, I-15121, Alessandria, Italy; (C)INFN, I-10125, Turin, Italy\\
% ...
$^{a}$ Also at the Novosibirsk State University, Novosibirsk, 630090, Russia\\
% ...
      }\end{center}
    \vspace{0.4cm}
  \end{small}
}

\begin{abstract}
\vspace{3mm}
We report the first observation of the Dalitz decay $\eta' \to \gamma e^+e^-$,
based on a data sample of 1.31 billion $J/\psi$ events collected with the BESIII detector.
The $\eta'$ mesons are produced via the  $J/\psi \to \gamma \eta^\prime$ decay process.
The ratio $\Gamma(\eta' \to \gamma e^+ e^-)/\Gamma(\eta'\to\gamma\gamma)$ is measured
to be $(2.13\pm0.09(\text{stat.})\pm0.07(\text{sys.}))\times10^{-2}$.
This corresponds to a branching fraction ${\cal B}(\eta' \to \gamma e^+e^-)=
(4.69 \pm0.20(\text{stat.})\pm0.23(\text{sys.}))\times10^{-4}$.
The transition form factor is extracted and different expressions are compared
to the measured dependence on the $e^+e^-$ invariant mass. The results are
consistent with the prediction of the Vector Meson Dominance model.
\vspace{5mm}
\end{abstract}

\maketitle

\section{Introduction}\label{Introduction}

Electromagnetic (EM) Dalitz decays of light pseudoscalar mesons,
$P\to \gamma l^+l^-$ ($P=\pi^0$, $\eta$, $\eta'$; $l=e, \mu$), play an
important role in revealing the structure of hadrons and the interaction
mechanism between photons and hadrons~\cite{Landsberg}.  If one assumes
point-like particles, the decay rates can be exactly calculated
by Quantum Electrodynamics (QED)~\cite{QED}.
Modifications to the QED decay rate due to the inner structure of
the mesons are encoded in the transition form factor (TFF) $F(q^2)$,
where $q$ is the momentum transferred to the lepton pair, and $q^2$ is the square of the invariant
mass of the lepton pair. A recent summary and discussion of this subject
can be found in Ref.~\cite{M1207}.

The knowledge of the TFF is also important in studies of the muon anomalous
magnetic moment, $a_\mu = (g_\mu -2)/2$, which is the most precise low-energy test of the
Standard Model (SM) and an important probe for new physics. The theoretical uncertainty on the
SM calculation of $a_\mu$ is dominated by hadronic corrections and therefore limited by
the accuracy of their determination~\cite{Blum:2013xva}.
In particular, the hadronic light-by-light (HLbL) scattering contribution to $a_\mu$ includes
two meson-photon-photon vertices that can be related to the form factors in
$P \to \gamma \gamma^* \to \gamma e^+e^-$ decays~\cite{Blum:2013xva}.
Thus, models describing these transitions should be tested as precisely as possible
to reduce the uncertainty in the SM prediction for $(g_\mu -2)/2$.

In this work, the Dalitz decay  $\eta^\prime\to\gamma e^+e^-$
is measured for the first time. The differential decay width,
normalized to the radiative decay width $\eta'\to\gamma\gamma$,
is~\cite{Landsberg}
\begin{eqnarray}\label{eq:decaywidth}
&&\frac{d\Gamma(\eta'\to\gamma l^+l^-)}{dq^2\Gamma(\eta'\to \gamma\gamma)}  \nonumber\\
&& ~~=\frac{2\alpha}{3\pi}\frac{1}{q^2}\sqrt{1-\frac{4m_l^2}{q^2}}\left(1+\frac{2m_l^2}{q^2}\right)\!\left(1-\frac{q^2}{m_{\eta'}^2}\right)^{\!\!3}\left|F(q^2)\right|^2 \nonumber\\
&& ~~=  [\mbox{QED}(q^2)] \times |F(q^2)|^2,
\end{eqnarray}
\noindent
where $m_{\eta'}$ and $m_l$ are the masses of the $\eta'$ meson and
the lepton, respectively; $\alpha$ is the fine structure constant;
and $[\mbox{QED}(q^2)]$ represents the calculable QED part
for a point-like meson. The TFF, $F(q^2)$, which
is described by phenomenological models, can be experimentally determined
from differences between the measured di-lepton invariant mass spectrum
and the QED calculation. In the Vector Meson
Dominance (VMD) model~\cite{JJS}, it is assumed that interactions between virtual
photon and hadrons are dominated by a superposition of neutral vector meson states.
One commonly used expression for the multi-pole form factor is~\cite{VMDFF}:

\begin{eqnarray}\label{eq:FFVMD}
F(q^2)=N\sum_V\frac{g_{\eta'\gamma V}}{2g_{V\gamma}}\cdot\frac{m^2_V}{m^2_V-q^2-i\Gamma_V m_V},
\end{eqnarray}
where $N$ is a normalization constant ensuring that $F(0)=1$;
$V=\rho, \omega, \phi$; $m_V$, $\Gamma_V$ are the masses and
widths of these vector mesons; and $g_{\eta'\gamma V}$
and $g_{V\gamma}$ are the corresponding coupling constants.

The parameter to be experimentally determined is the slope of the
form factor $b$, which is related to the effective virtual
vector meson mass $\Lambda$ by
\begin{eqnarray}\label{eq:slope}
 b=\frac{dF}{dq^2}\Big|_{q^2=0} =  \Lambda^{-2}
\end{eqnarray}
In experiments, the single-pole form factor is generally used to extract
the slope of the form factor. For the case of the $\eta'$, the pole is
expected to lie within the kinematic boundaries of the decay. The square of the
form factor is described by
\begin{eqnarray}\label{FF2}
   |F(q^2)|^2=\frac{\Lambda^2(\Lambda^2+\gamma^2)}{(\Lambda^2-q^2)^2+\Lambda^2\gamma^2}
\end{eqnarray}
where the parameters $\Lambda$ and $\gamma$ correspond to the mass
and width of the Breit-Wigner shape for the effective contributing vector meson.
To a first approximation, $\Lambda\approx M_\rho\approx$ 0.7~GeV
and $\gamma\approx \Gamma_{\rho}\approx$ 0.12~GeV.

For the $\etap$ Dalitz decay, only the process $\etap \to \gamma \mu^+\mu^-$
has been observed and the slope of the form factor was measured to be
$b_{\eta'} = (1.7\pm0.4)$~GeV$^{-2}$~\cite{Landsberg, etapGmu}. To date,
the process $\eta' \to \gamma e^+e^-$ has not been observed yet. The most stringent upper
limit on the ratio of decay widths $\Gamma(\eta'\to\gamma e^+e^-)/\Gamma(\eta'\to\gamma\gamma)$
is $4.1\times10^{-2}$ at the 90\% confidence level (CL) from the CLEO Collaboration~\cite{CLEO},
which is above the predicted value of $(2.06\pm0.02)\times10^{-2}$
from the modified VMD model~ \cite{Petri}.

In the VMD model, the TFF slope is expected to be $b_{\eta'} = 1.45$~GeV$^{-2}$~\cite{betap1, betap2},
while for chiral perturbation theory it is
$b_{\eta'}=1.60$~GeV$^{-2}$~\cite{chP}. A recent calculation
based on a dispersion integral gives
$b_{\eta'} = 1.53^{+0.15}_{-0.08}$~GeV$^{-2}$~\cite{M1309}.

We report the first observation of the $\eta'\to \gamma e^+e^-$ decay
and the extraction of the TFF. The source of the $\eta'$ mesons
are radiative $J/\psi\to\gamma\eta'$ decays in a sample of 1.31 billion
$J/\psi$ events ($2.25 \times 10^8$ events were taken in 2009~\cite{Jpsi09} and $1.09 \times 10^9$ in 2012)~\cite{Jpsi12}
collected by the BESIII~\cite{BESIII}
at the BEPCII $e^+e^-$ collider.  The $\eta'\to\gamma\gamma$ decay events
in the same data sample are used for normalization.

\section{The BESIII experiment and Monte Carlo simulation} \label{Detector}
BEPCII is a double-ring multi-bunch $e^+e^-$ collider running
in the tau-charm energy region. The BESIII detector,
described in detail in Ref.~\cite{BESIII},
has a geometrical acceptance of 93\% of 4$\pi$ solid angle. It
consists of a drift chamber (MDC), a time-of-flight (TOF) system,
and an electromagnetic calorimeter (EMC), all enclosed in a super-conducting solenoid
with 1.0~T (0.9~T in 2012) magnetic field. The small-cell helium based MDC
provides the tracking of charged particle and ionization
energy loss ($dE/dx$) measurement. The single cell position
resolution is 130 $\mu$m and the transverse momentum resolution
is 0.5\% at 1 GeV/c. The TOF system for particle identification (PID)
is made of plastic scintillators. It has 80~ps time resolution in
the barrel, and 110~ps in the end caps. The EMC is made of 6240 CsI~(Tl)
crystals. The energy resolution is 2.5\% in the barrel and 5\% in
the end caps for 1.0~GeV photons. Outside the solenoid, a muon chamber system made
of 1272~m$^2$ resistive plate chambers detects muon tracks with
momenta greater than 0.5~GeV/c.

The \textsc{geant4}-based~\cite{GEANT4} simulation
software BOOST includes the description of geometry and material of the
BESIII detector, the detector response and digitization models, and also
tracks the detector running conditions and performance.
A Monte Carlo (MC) simulated sample of 1.2 billion $J/\psi$ inclusive decays is used to study potential backgrounds.
The production of the $J/\psi$ resonance
is simulated by the MC event generator \textsc{kkmc}~\cite{KKMC}; the known
decay modes are generated by \textsc{evtgen}~\cite{GEN, bes3gen} with
branching fractions set at the world average values~\cite{PDG}, while unknown
decays are generated by \textsc{lundcharm}~\cite{LUND}.
The \textsc{evtgen} package is used to generate $J/\psi\to \gamma\eta'$, $\eta'\to\gamma e^+e^-$
and $\eta'\to\gamma \gamma$ events. The decay $J/\psi\to\gamma\eta'$ is generated with an
angular distribution of $1+\cos^2\theta_{\gamma}$, where $\theta_{\gamma}$ is the radiative photon
angle relative to the positron beam direction in the $J/\psi$ rest frame.
In generating $\eta'\to\gamma e^+e^-$, the TFF is parameterized by
the multi-pole VMD model in Eq.(\ref{eq:FFVMD}) with the parameters
taken from Ref.~\cite{Landsberg}.

\section{Signal selection: $J/\psi\to\gamma \eta', \eta' \to \gamma e^+e^-$} \label{gammaee}

Charged tracks are reconstructed from hits registered in the MDC.
Only tracks with $|\cos\theta| < 0.93$ are retained, where $\theta$ is the polar angle with respect
to the beam axis. The tracks are required to pass within 10~cm of
the center of the interaction region in the beam direction ($Z$ axis) and within 1~cm in the plane
perpendicular to the beam. Event candidates are required to
have two well reconstructed charged tracks with net
charge zero. For electron identification, information from $dE/dx$ and TOF
is combined to compute probabilities for the electron ($\rm{Prob}(e)$)
and pion ($\rm{Prob}(\pi)$) hypothesis.
To separate electrons from pions, we require $\rm{Prob}(e)/(\rm{Prob}(e)+\rm{Prob}(\pi))>0.95$.
Final states with kaons cannot contribute to
the background because of the limited phase space.

Electromagnetic showers are reconstructed from clusters of energy deposits
in the EMC. The photon candidate showers must have a minimum energy of 25~MeV
in the barrel region $(|\cos\theta | < 0.80)$ and  50~MeV
in the end cap region $(0.86 < |\cos\theta| < 0.92)$. Showers in the
region between the barrel and the end caps are poorly
measured and excluded from the analysis. To exclude charged-particle induced activities,
the showers are required to be separated from the extrapolated positions
of any charged track by at least $10^{\circ}$. In addition, cluster timing requirements are used
to suppress electronic noise and unrelated energy deposits.

A vertex fit is performed on the
electron and positron tracks, and a loose $\chi^2$ requirement is applied to ensure
that they come from a common vertex.
To improve resolution and reduce background, a four-constraint (4C)
kinematic fit is performed to the $\gamma\gamma e^+e^-$ hypothesis that constrains
the total four-momentum of the detected particles to be equal to the initial four-momentum of
the colliding beams. For events with more than two photon candidates, the combination
with the smallest $\chi^2_{\rm{4C}}$ is selected. Only events with $\chi^2_{\rm{4C}}<$100
are retained.

For the $J/\psi \to \gamma \eta', \eta'\to\gamma e^+e^-$ signal channel,
the largest background comes from QED processes  and $J/\psi\to e^+e^-\gamma\gamma$ decays.
For these channels, the combination of the $e^+e^-$ with any final-state photon
produces a smooth $M(\gamma e^+e^-$) distribution.
The QED background mainly comes from $e^+e^-\to e^+e^-\gamma\gamma$
and $e^+e^-\to 3\gamma$ events in which one $\gamma$ converts into
an $e^+e^-$ pair. These are studied using a $e^+e^-$ collision data sample of
2.92~fb$^{-1}$ taken at $\sqrt{s}=3.773$~GeV~\cite{3773}, which
is dominated by QED processes. For those processes, most of the photons
have low energy and are at small angles relative to the incoming electron
or positron beam directions. To reduce this background, the energy of the low-energy photon
is further required to be higher than 200~MeV, and the angle between the photon
and the electron or positron initial direction in the final states is required to be larger than $10^{\circ}$.

\begin{figure}[htbp]
    \centering
        \includegraphics[width=4.3cm]{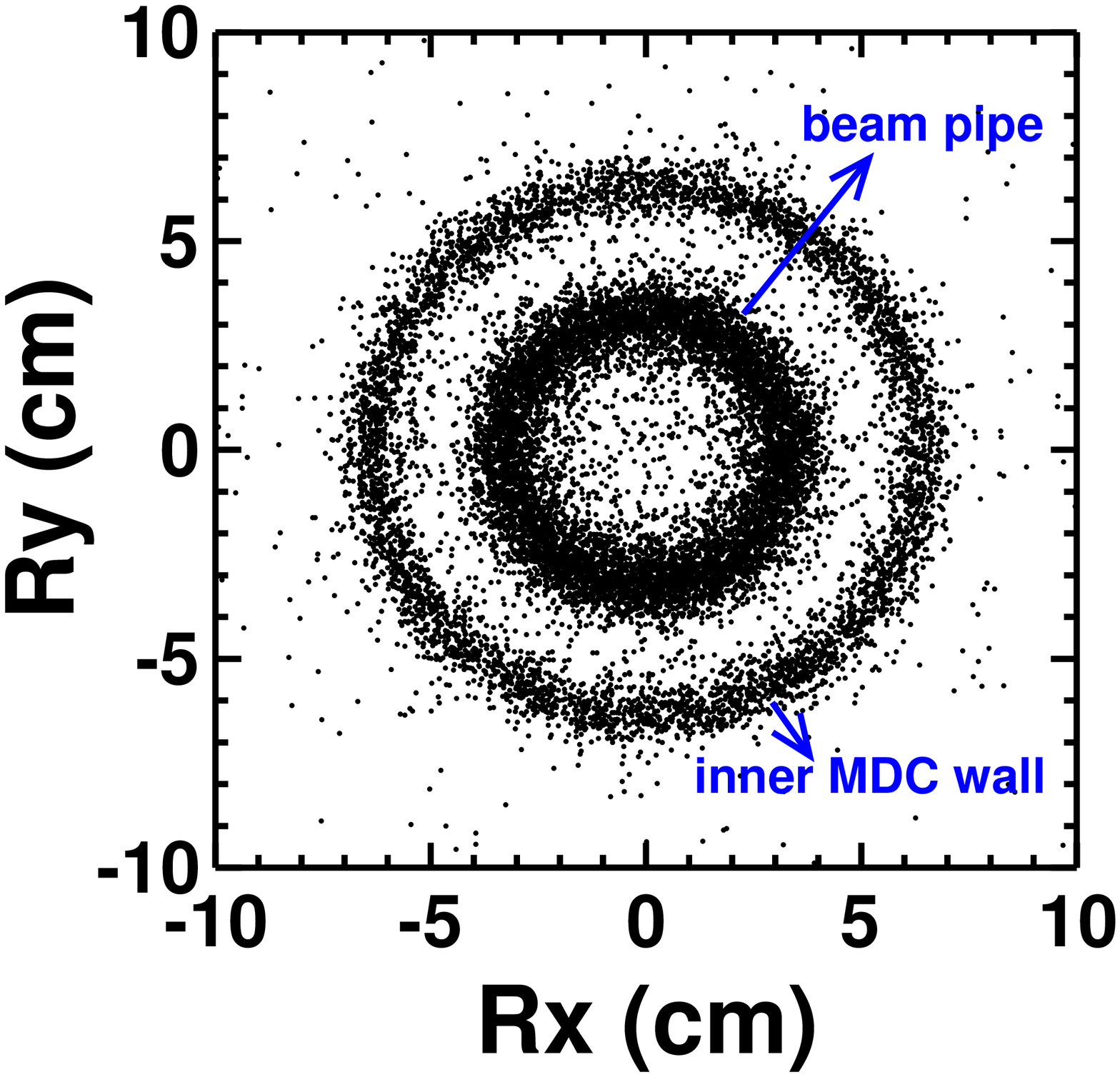}\put(-98,90){\bf \large~(a)}
        \includegraphics[width=4.3cm]{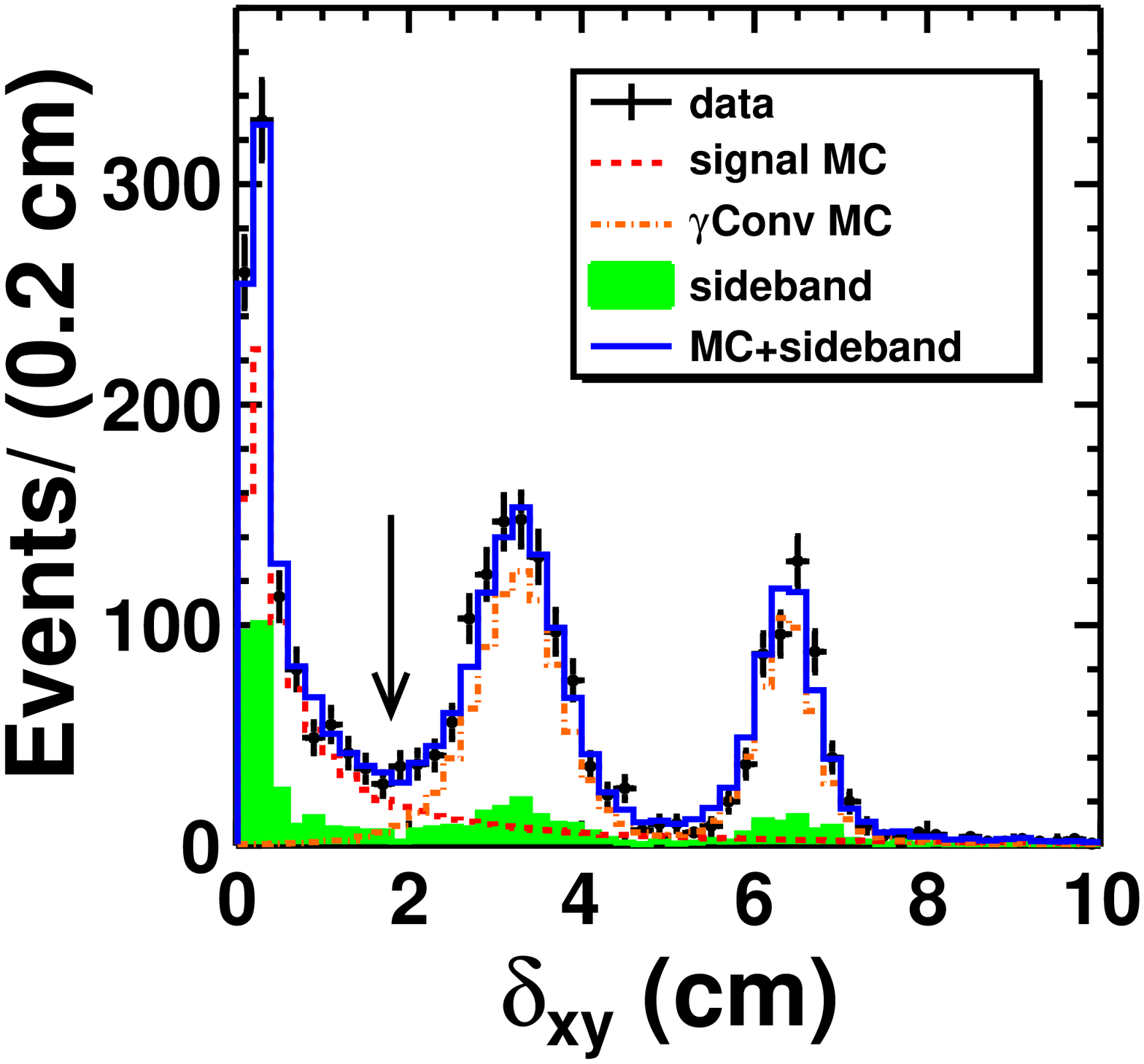}\put(-95,90){\bf \large~(b)}
        \caption{
        Electron-positron vertex position distribution: (a) scatter plot of $R_{y}$ versus $R_{x}$ for MC-simulated
        $J/\psi \to \gamma\etap$, $\etap \to \gamma\gamma$ events. (b) $\delta_{xy}$ distributions. The (black)
        crosses are data. The (red) dashed line shows the MC-simulated $J/\psi \to\gamma\etap$,
        $\etap \to \gamma e^+e^-$ signal events. The (orange) dotted-dashed histogram shows the background from
        $\gamma$-conversion events. The (green) shaded area is estimated from the $\etap$ mass sideband. The
        (blue) line is the sum of MC and the sideband estimate. In (b), the solid arrow indicates the requirement
        on $\delta_{xy}$.}
    \label{rxy}
\end{figure}

The primary peaking background comes from the decay
$J/\psi\to\gamma\eta',\eta' \to \gamma\gamma$  followed by a $\gamma$ conversion in the material in front of
the MDC, including the beam pipe and the inner wall of the
MDC.  The distance from the reconstructed vertex
point of the electron-positron pair to the $z$ axis, defined as
$\delta_{xy}=\sqrt{R^2_x+R^2_y}$, is used to distinguish $\gamma$
conversion events from signal events~\cite{GammaConv}, where $R_x$ and $R_y$ are the distances in
the $x$ and $y$ directions, respectively.
A scatter plot of $R_y$ versus $R_x$ is shown in Fig.~\ref{rxy}~(a) for MC-simulated
$J/\psi\to\gamma\eta',\eta' \to \gamma\gamma$ decays, in which one of the photons
undergoes conversion to an $e^+e^-$ pair.
As indicated in Fig.~\ref{rxy}~(a), the inner circle matches the position
of the beam pipe, while the outer circle corresponds to the position of the inner
wall of the MDC. Figure~\ref{rxy}~(b) shows the $\delta_{xy}$
distributions for the MC simulated $J/\psi \to \gamma\eta', \eta'\to\gamma e^+e^-$,
$J/\psi \to \gamma\eta', \eta'\to\gamma \gamma$ events, together with the selected data events and
events from the $\eta'$ mass sideband.
The two peaks above 2.0~cm correspond to the photon conversion of the $\gamma$ from
$J/\psi\to\gamma\eta',\eta' \to \gamma\gamma$ events, while the events near
$\delta_{xy}=0$~cm originate from the interaction point. We require $\delta_{xy}<2$ cm to suppress the
photon-conversion background, which retains about 80\%
of the signal events while removing about 98\% of the photon-conversion
events.
After all selections, the normalized number of expected peaking background events from
$J/\psi\to\gamma\eta',\eta' \to \gamma\gamma$ is $42.7\pm8.0$, where the error is dominantly from
the difference in selection efficiencies for the $\gamma$ conversion events between data and MC.

Another possible source of peaking background is $J/\psi\to\gamma\eta', \eta'\to\gamma\pi^+\pi^-$,
where the two pions are misidentified as an $e^+e^-$ pair.
The momenta of these pions are most below 200~MeV/$c$.
An exclusive MC sample that includes coherent contributions from $\rho$, $\omega$, and the box anomaly
in the decay $\eta'\to\gamma\pi^+\pi^-$~\cite{GammaRho} is used to study this background.  We find that
the kinematic fit to the electron-positron hypothesis shifts the spectrum away from the $\eta'$ mass and,
thus, the resulting M($\gamma e^+e^-$) distribution does not peak at the $\eta'$ mass value.
The normalized number of events from this background source after all
selections is $9.7\pm0.4$, which is negligible compared
to the non-peaking background from $e^+e^-\to e^+e^-\gamma\gamma$.

\begin{figure}[hbtp]
  \centering
  \includegraphics[width=0.8\linewidth]{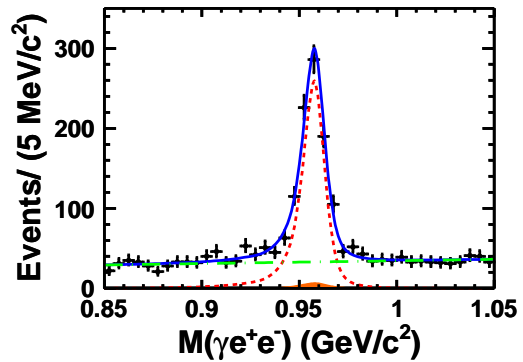}
  \caption{Invariant $\gamma e^+e^-$ mass distribution for the
    selected signal events. The (black) crosses
           are the data, the (red) dashed line represents the signal, the (green) dot-dashed curve shows
           the non-peaking background shapes, the (orange) shaded component is the shape of the
           $J/\psi \to \gamma \eta', \eta'\to\gamma \gamma$ peaking
           background events. The total fit result is shown
           as the (blue) solid line.}
  \label{etap:fit}
\end{figure}

The combination of $\gamma e^+e^-$ with invariant mass closest to $m_{\eta'}$ is taken to reconstruct the $\eta'$.
 The resulting $M(\gamma e^+e^-)$ distribution after the selection criteria is shown
in Fig.~\ref{etap:fit} and exhibits a clear peak at the $\eta'$ mass.
An unbinned extended maximum likelihood (ML) fit is performed to
determine the signal yield. The signal probability density function (PDF)
is represented by the signal MC shape. The shape for the non-peaking
background is described by a first-order Chebychev polynomial.
The background yield and its PDF parameters are allowed to vary in the fit.
The peaking background from the $\gamma$-conversion of
$J/\psi \to \gamma \eta', \eta'\to\gamma \gamma$ decay is obtained
from the MC-simulated shape with the yield fixed as described before. The fitting range is 0.85$-$1.05~GeV/$c^2$.
The net signal yield and the detection efficiency are summarized in Table~\ref{yields}.

\begin{table}[hbtp]
  \caption{Number of observed signal events, $N_{\eta'\to\gamma e^+e^-}$ ($N_{\eta'\to\gamma \gamma}$),
           and detection efficiency,  $\epsilon_{\eta'\to\gamma e^+e^-}$ ($\epsilon_{\eta'\to\gamma \gamma}$)
           for $J/\psi\to\gamma\eta', \eta'\to\gamma e^+e^-$ ($J/\psi\to\gamma\eta', \eta'\to\gamma \gamma$).
           The uncertainties are statistical only.}
  \label{yields}\footnotesize
  \begin{center}
     \renewcommand{\arraystretch}{1.8}
     \begin{tabular}{c|cc}
        \hline\hline
                          & \hspace{0.5cm} $\eta'\to\gamma e^+e^-$ \hspace{0.5cm} &  \hspace{0.5cm} $\eta'\to\gamma\gamma$   \hspace{0.5cm} \\ \hline
          $N_{\eta'\to\gamma e^+e^-}$ ($N_{\eta'\to\gamma \gamma}$)            &   $864\pm36$          &  $70846\pm292$      \\

        $\epsilon_{\eta'\to\gamma e^+e^-}$ ($\epsilon_{\eta'\to\gamma \gamma}$)      &      24.5\%             & 42.8\%     \\

        \hline\hline
      \end{tabular}
  \end{center}
\end{table}

\section{\boldmath Normalization channel: $J/\psi\to\gamma \eta', \eta' \to \gamma \gamma$} \label{gammagamma}

The decay $J/\psi\to\gamma \eta', \eta' \to \gamma \gamma$ is studied using the same data set, and we
quote our result in terms of the ratio $\Gamma(\eta'\to\gamma e^+e^-)/\Gamma(\eta'\to \gamma\gamma)$.
In this ratio the uncertainties due to the total number of $J/\psi$
events and the branching fraction for
$J/\psi\to\gamma\eta'$ cancel, and the uncertainty due to the photon detection efficiency partially cancels.

Events with zero charged particles and at least three photon candidates are selected with
the same requirements that are used
for the signal events.  A 4C kinematic fit is performed to the $J/\psi \to\gamma\gamma \gamma$ hypothesis.
For events with more than three photons, the combination with the smallest \chisq{4C}
is selected. The \chisq{4C} is required to be less than 100.  The two photon
combination with invariant mass $M(\gamma\gamma)$ closest to $m_{\eta'}$ is taken as from
the $\eta'$ decay.

Detailed MC studies indicate that no peaking background remains after all the selection criteria.
The non-peaking background mainly comes from the
 continuum process $e^+e^-\to\gamma\gamma\gamma$ and
 $J/\psi\to\gamma\pi^0\pi^0$ decays.  The latter source
involves intermediate states such as the $f_0(1500), f_0(1710), f_0(2020), f_2(1270), f_4(2050)$.
Because the $\eta'$ decays isotropically,
the angular distribution of photons from the $\eta'$ decays is flat in
$\cos\theta_{\text{decay}}$, where $\theta_{\text{decay}}$ is the angle of the decay photon
in the $\eta^\prime$  helicity frame. In contrast, background events from QED continuum processes
and $J/\psi\to\gamma\pi^0\pi^0$ decays tend to accumulate near $\cos\theta_{\text{decay}}=\pm 1$.
We suppress these non-peaking backgrounds by requiring $|\cos\theta_{\text{decay}}|<0.8$.

\begin{figure}[hbtp]
  \centering
    \includegraphics[width=0.8\linewidth]{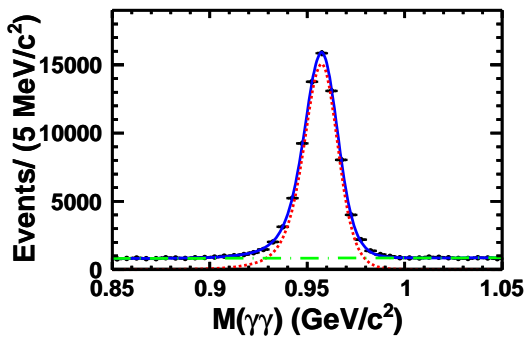}
  \caption{Invariant mass distribution of $M(\gamma \gamma)$ for the
    selected events in the normalization channel.
           The (black) crosses are data, the (red) dashed curves
           represents the $\eta' \to \gamma \gamma$ signal, the
           (green) dot-dashed curve shows the non-peaking background. The fit result is
           shown as the (blue) solid curve.}
  \label{etapgg:fit}
\end{figure}

The $M(\gamma\gamma)$ distribution for events that survive the selection requirements
is shown in Fig.~\ref{etapgg:fit}. An unbinned ML fit
is performed to obtain the yield of $J/\psi \to \gamma \eta', \eta'\to \gamma \gamma$.
The PDF used to represent the signal is taken from the MC, and the PDF for the non-peaking background is
a first-order Chebychev polynomial with coefficients determined from the fit. The resulting
signal yield and the MC-determined detection efficiency are summarized in Table~\ref{yields}.

\begin{table}[hbtp]
  \caption{Summary of relative systematic uncertainties for the
    determination of the ratio
            $\frac{\Gamma(\eta' \to \gamma e^+
              e^-)}{\Gamma(\eta'\to\gamma\gamma)}$. The last row is
            the uncertainty associated with the
           $J/\psi \to \gamma \eta^\prime$, $\eta' \to \gamma \gamma$ normalization sample. }
  \label{sys:total}
  \begin{center}
     \renewcommand{\arraystretch}{1.5}
     \begin{tabular}{c|c}
        \hline\hline
       Sources  & Systematic uncertainties(in \%)   \\ \hline

        MDC tracking                     & 0.6      \\

        PID                               & 0.6      \\

        Photon detection                  & 1.0        \\

        Veto of gamma conversion          & 1.8        \\

        4C kinematic fit                  & 1.0           \\

        Form factor uncertainty          & 1.6       \\

        Fit range \& Bkg shape           & 0.9      \\

        Uncertainty of $N_{\eta'\to\gamma\gamma}$  & 1.2  \\

         \hline
        Total       &    3.3   \\
        \hline\hline
      \end{tabular}
  \end{center}
\end{table}

\section{Systematic uncertainties in the relative decay width}\label{systemtic:section}

Table~\ref{sys:total} lists all sources of systematic
uncertainties associated with the measurement of the ratio
$\Gamma(\eta'\to\gamma e^+e^-)/\Gamma(\eta'\to \gamma\gamma)$.
Most systematic uncertainties are determined from
comparisons of low-background, high-statistics data samples with results
from MC simulations.

The electron and positron tracking and PID efficiencies are determined
using a sample of radiative Bhabha $e^+e^-\rightarrow \gamma e^+e^-$
(including $J/\psi\to\gamma e^+ e^-$) events collected
at the $J/\psi$ energy.  Differences in tracking and PID
efficiencies between data and MC simulation are determined for every bin of a
two-dimensional distribution of the momentum versus polar angle of the lepton tracks.
These are used to determine an overall weighted difference per track of
$(1.1\pm0.3)$\% for the tracking efficiency and $(1.9\pm0.3)$\% for the PID efficiency.
The MC efficiency is corrected for these differences, and the uncertainties of the correction
coefficients are assigned as the systematic uncertainties associated with the lepton tracking and PID
efficiencies.

The systematic uncertainty associated with the photon detection efficiency is
studied using three different methods, as described in Ref.~\cite{photon}.
The three methods provide consistent results for the photon efficiency uncertainty to be
1\% per photon. Because the systematic uncertainty from the
radiative photon and one photon from the $\eta'$ cancel in the ratio, the
total systematic uncertainty from photon detection is 1\%.

In the analysis, the peaking background from $J/\psi\to\gamma\eta', \eta'\to\gamma \gamma$
$\gamma$-conversion events is suppressed by the requirement $\delta_{xy}<2$~cm. To estimate
the systematic uncertainty associated with this requirement, we use a sample of $J/\psi \to
\pi^+\pi^-\pi^0, \pi^0 \to \gamma e^+e^-$ that includes both
$\pi^0$ Dalitz decays and $\pi^0 \to \gamma \gamma$ decays with one
photon externally converted to an electron-positron pair.  The data-MC difference
of 1.8\% for these events is considered as the systematic uncertainty for our $\gamma$-conversion
veto requirement on $\delta_{xy}$.

A systematic uncertainty associated with the kinematic fit will occur if the track-helix parameters
for data and MC simulated events are not consistent. Following the procedure described in
Ref.~\cite{GuoYuping}, we use the $J/\psi \to \pi^+\pi^-\pi^0, \pi^0 \to \gamma e^+e^-$
decay as a control sample to extract the correction factors from the
pull distributions of the track helix parameters. The 1\% difference between the efficiencies
with and without helix parameter corrections is taken as the
systematic uncertainty.

To estimate the systematic uncertainty due to the efficiency dependence on the choice of form-factor
parameterizations, signal MC events are also generated using a single-pole VMD model, shown in Eq.~(\ref{FF2}),
with $\Lambda = (0.79\pm0.05)$~GeV and $\gamma = (0.13\pm0.06)$~GeV, which are taken from
the fitted results described below in Section~\ref{FFM}. The relative difference in
the detection efficiency compared to that of the multi-pole model is taken
as the uncertainty associated with the form-factor parameterization.

In the fit to the $\gamma e^+e^-$ mass distribution, the signal PDF is fixed to the signal MC shape.
An alternative fit is performed by using a convolution of a MC signal shape with a Gaussian function
that is used to describe the MC-data difference due to the resolution.
The fitted width of the Gaussian is $(0.39\pm 0.19)$~MeV, and the fit yields $863.8\pm 36.0$ signal events.
The difference from the nominal fit is negligible.  Finally, the uncertainty due to the non-peaking
background shape is estimated by varying the PDF shape and fitting range in the ML fit.
The changes in yields for these variations give systematic uncertainties due to these backgrounds.

The systematic uncertainty in the measurement of $J/\psi\to\gamma\eta', \eta'\to\gamma\gamma$
associated with the uncertainty from the kinematic fit is estimated using
a control sample of $e^+e^-\to\gamma\gamma\gamma$ at 3.650~GeV~\cite{3Gamma} and found
to be less than 1\%. The uncertainty for this channel due to background is estimated
to be less than 0.3\% from variations in the PDF shape and fitting
range.  The uncertainty from the requirement $|\cos\theta_{\text{decay}}|<0.8$ is 0.4\%.
When combined with the 0.4\% statistical uncertainty, the total uncertainty
associated with $N_{\eta'\to\gamma\gamma}$ is 1.2\%.

Assuming all systematic uncertainties in Table~\ref{sys:total} are independent,
the total systematic uncertainty, obtained from their quadratic sum, is 3.3\%.

\section{Relative decay width}

The ratio $\Gamma(\eta'\to\gamma e^+e^-)/\Gamma(\eta'\to\gamma\gamma)$
is determined using the following formula:
\begin{eqnarray}
  \label{etapGEE:Br}
  \frac{\Gamma(\etap \to \gamma e^+ e^-)}{\Gamma(\etap\to\gamma\gamma)}&=&\frac{N_{\etap\to\gamma e^+e^-}}{N_{\etap\to\gamma\gamma}}\cdot\frac{\epsilon_{\etap\to\gamma\gamma}}{\epsilon_{\etap\to\gamma e^+e^-}},
\end{eqnarray}
where  $N_{\eta'\to\gamma e^+e^-}$ ($N_{\eta'\to\gamma \gamma}$) and $\epsilon_{\eta'\to\gamma e^+e^-}$
($\epsilon_{\eta'\to\gamma \gamma}$) are the number of observed signal events and the detection efficiency,
respectively, for $J/\psi\to\gamma\eta', \eta'\to\gamma e^+e^-$ ($J/\psi\to\gamma\eta',
\eta'\to\gamma \gamma$) decays, as listed in Table~\ref{yields}.
The result is
\begin{equation}
 \frac{\Gamma(\etap \to \gamma e^+ e^-)}{\Gamma(\etap\to\gamma\gamma)} =
(2.13\pm0.09(\text{stat.})\pm0.07(\text{sys.}))\times10^{-2}.
\end{equation}
Using the  $\eta'\to \gamma \gamma$ branching fraction value listed in PDG~\cite{PDG}, we obtain
the first measurement of the $\eta' \to \gamma e^+e^-$ branching fraction of
\begin{equation}
{\cal B}(\eta' \to \gamma e^+e^-)= (4.69 \pm0.20(\text{stat.})\pm0.23(\text{sys.}))\times10^{-4}.
\end{equation}

\begin{figure*}[hbtp]
  \centering
    \subfigure{
    \includegraphics[width=0.3\linewidth]{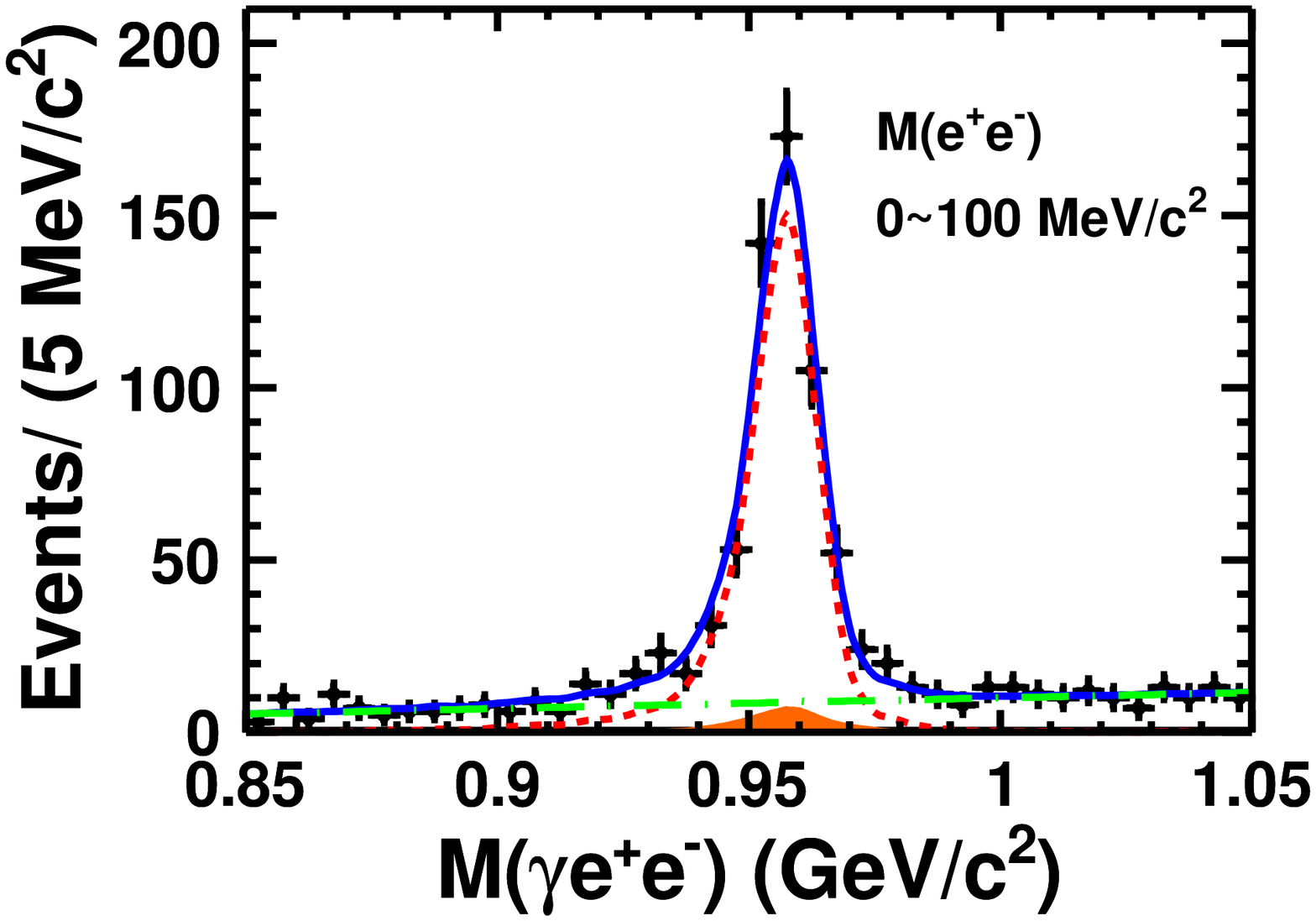}}\put(-120,65)
  \subfigure{
    \includegraphics[width=0.3\linewidth]{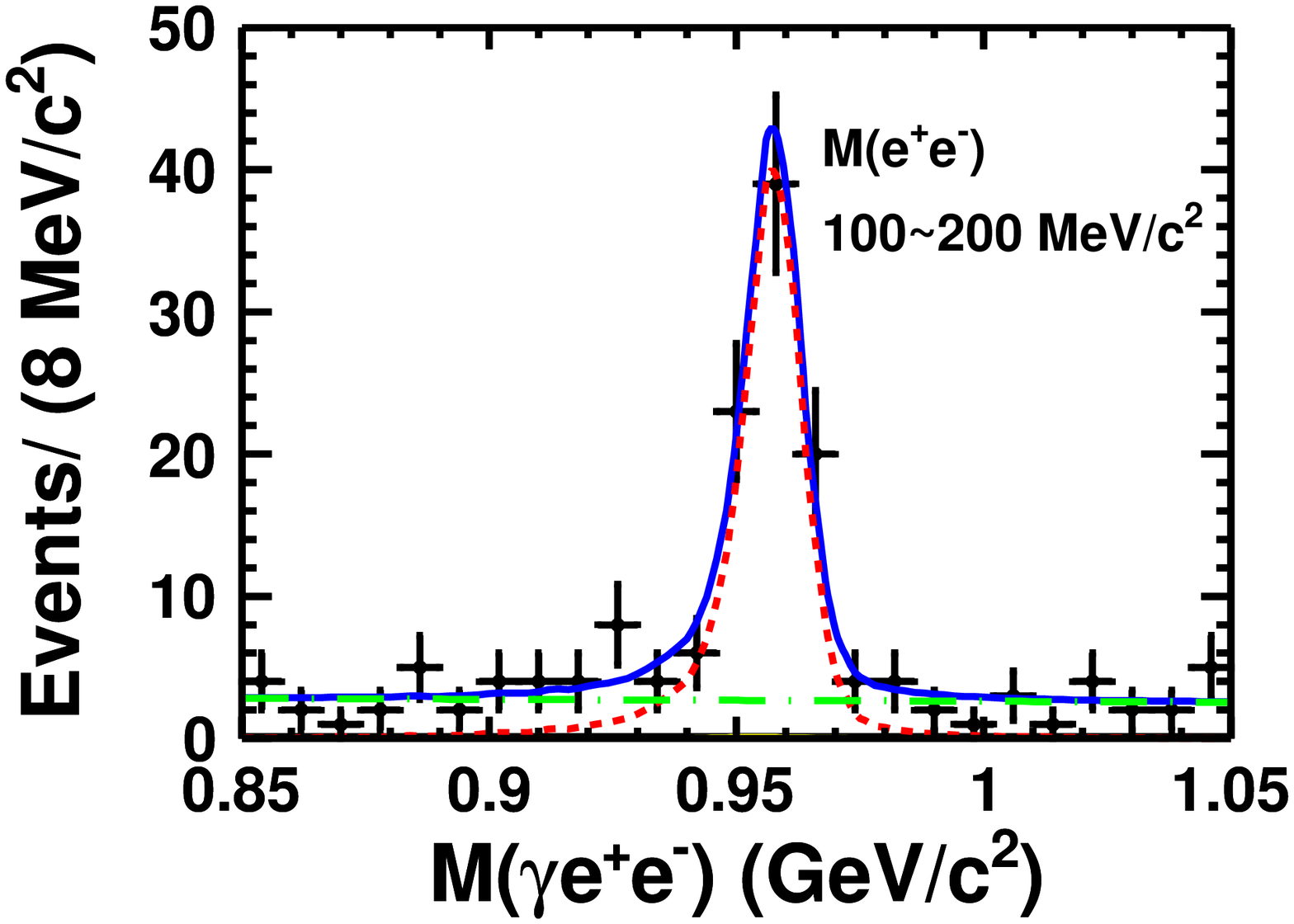}}\put(-120,65)
  \subfigure{
    \includegraphics[width=0.3\linewidth]{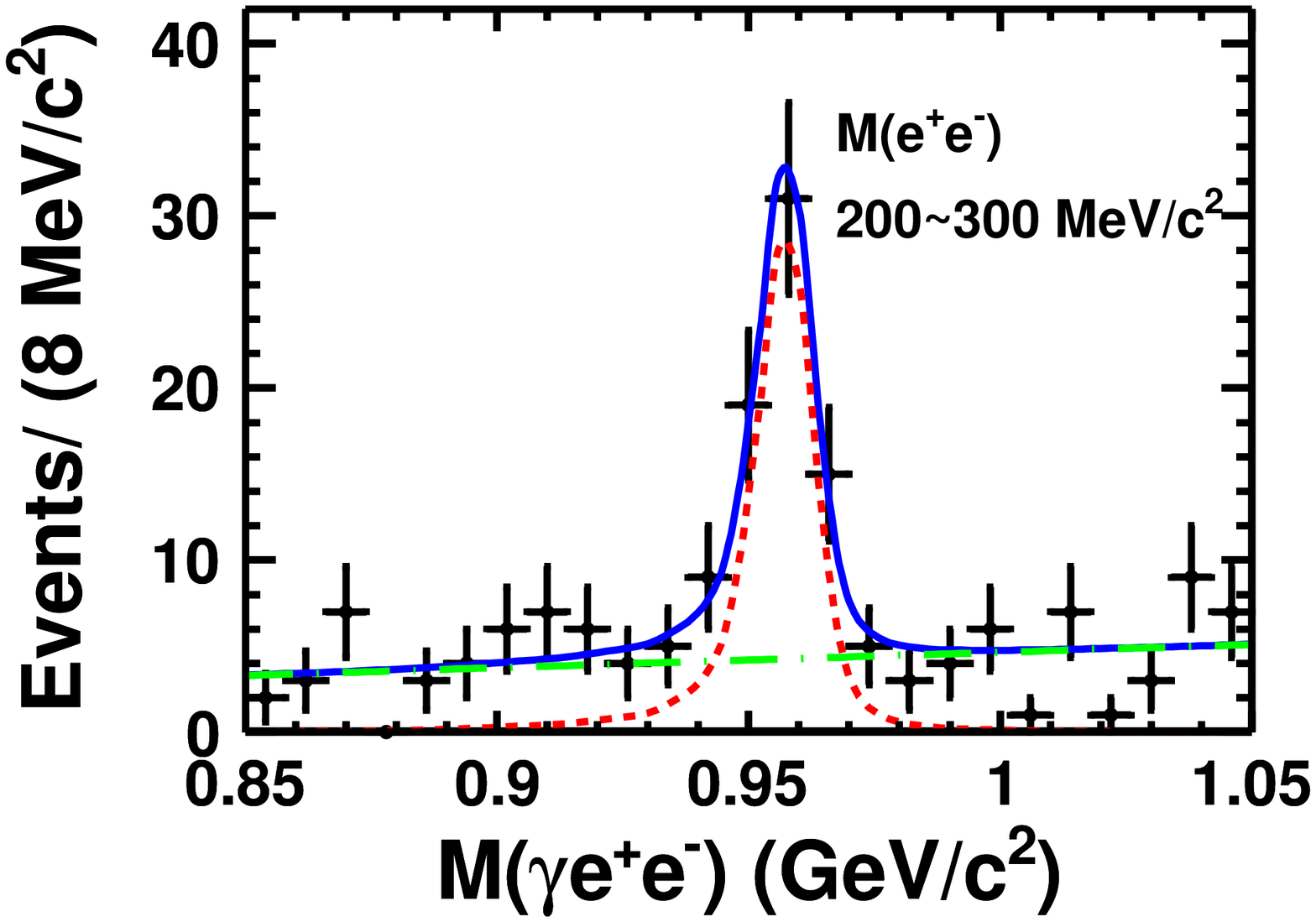}}\put(-120,65){}
  \\
  \subfigure{
    \includegraphics[width=0.3\linewidth]{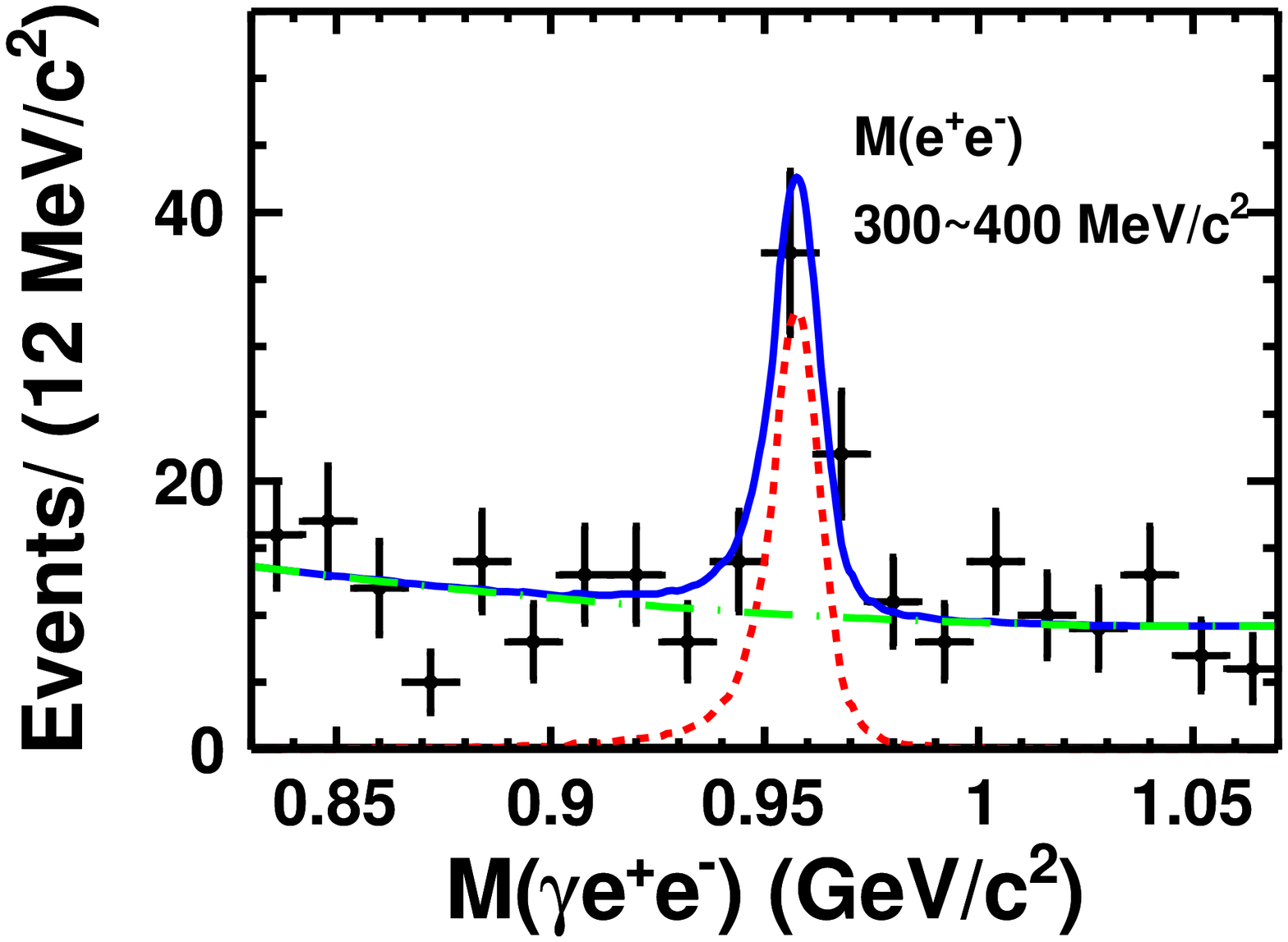}}\put(-120,65)
  \subfigure{
    \includegraphics[width=0.3\linewidth]{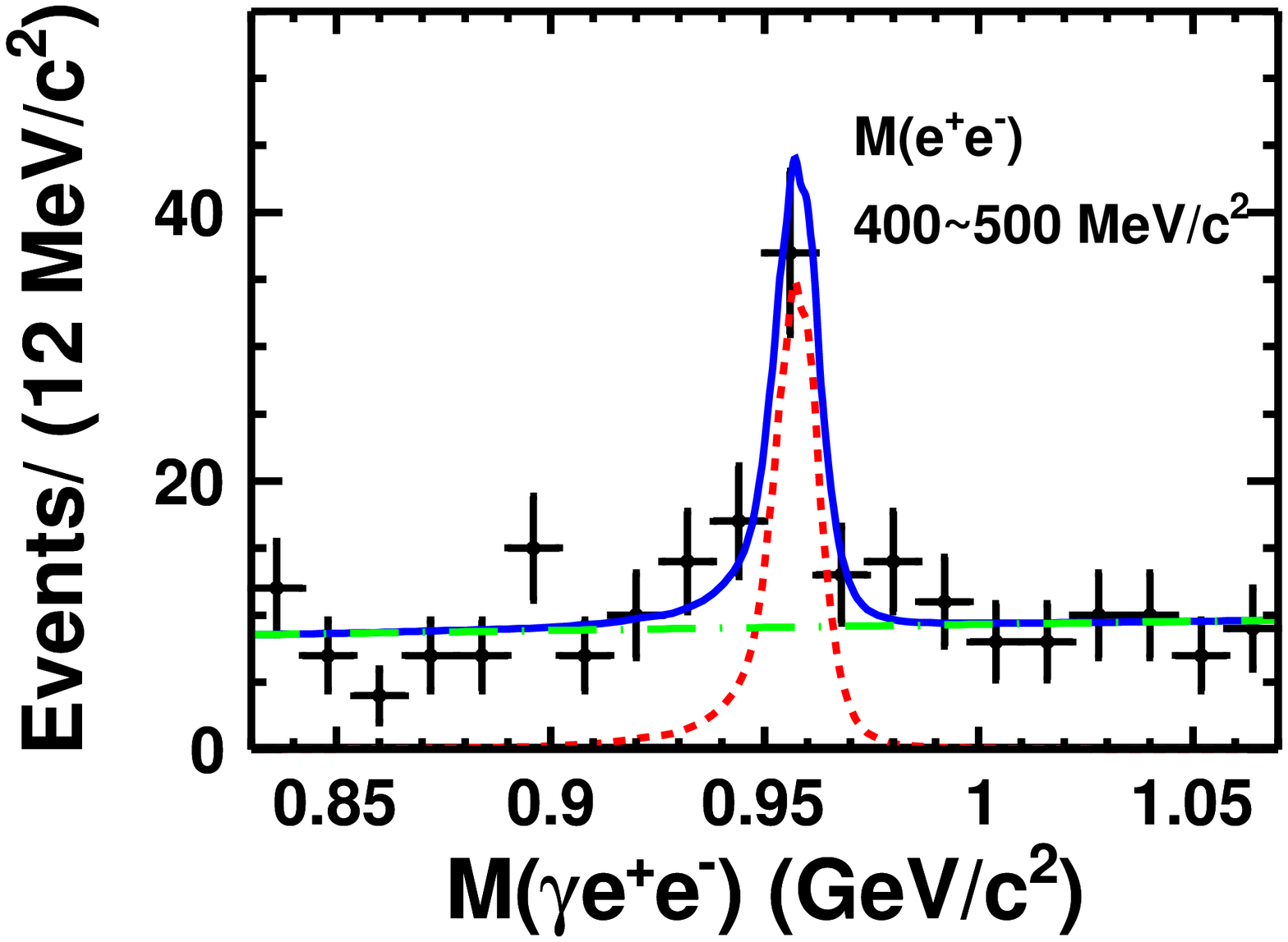}}\put(-120,65){ }
  \\
   \subfigure{
    \includegraphics[width=0.3\linewidth]{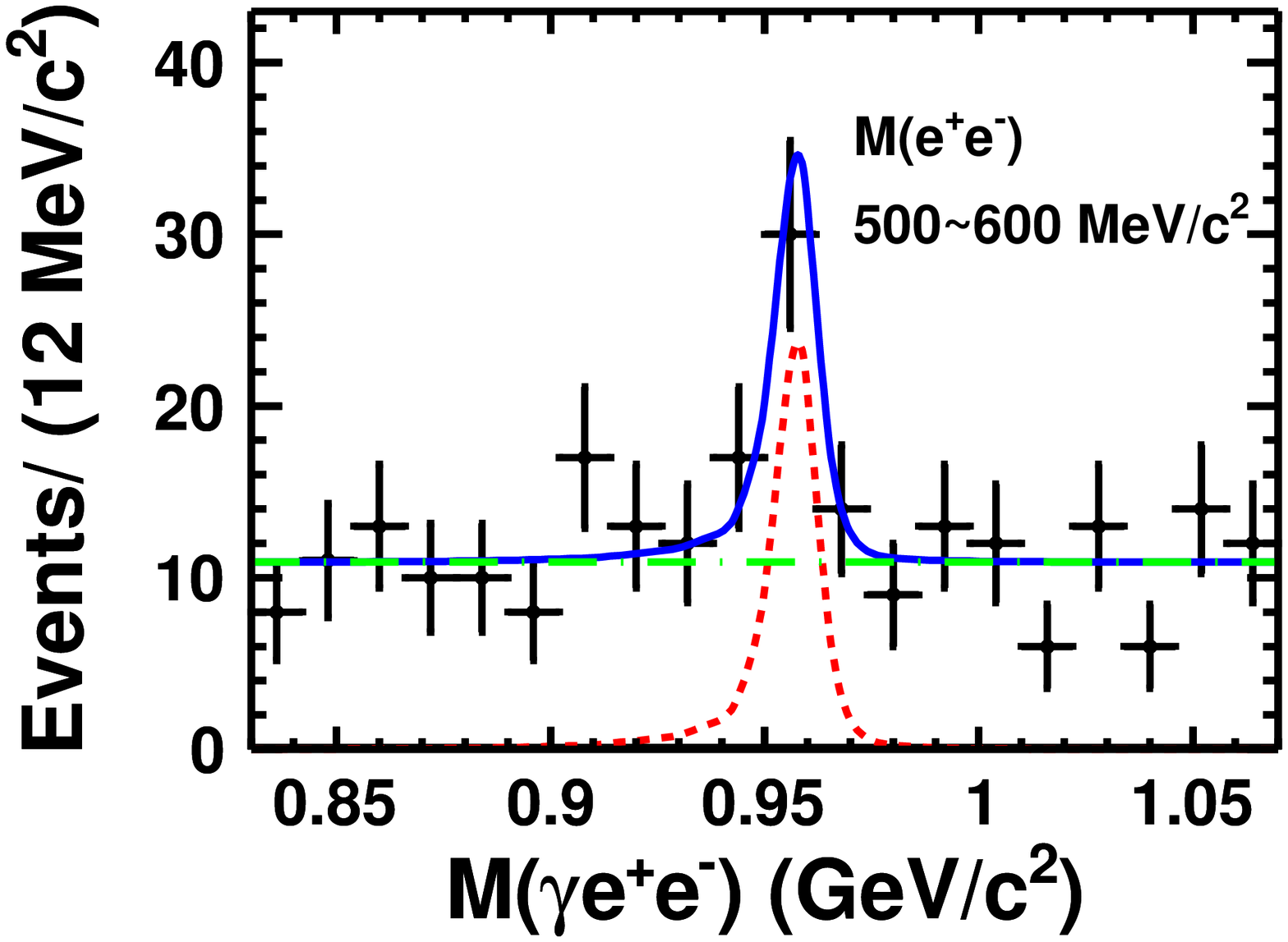}}\put(-120,65)
  \subfigure{
    \includegraphics[width=0.3\linewidth]{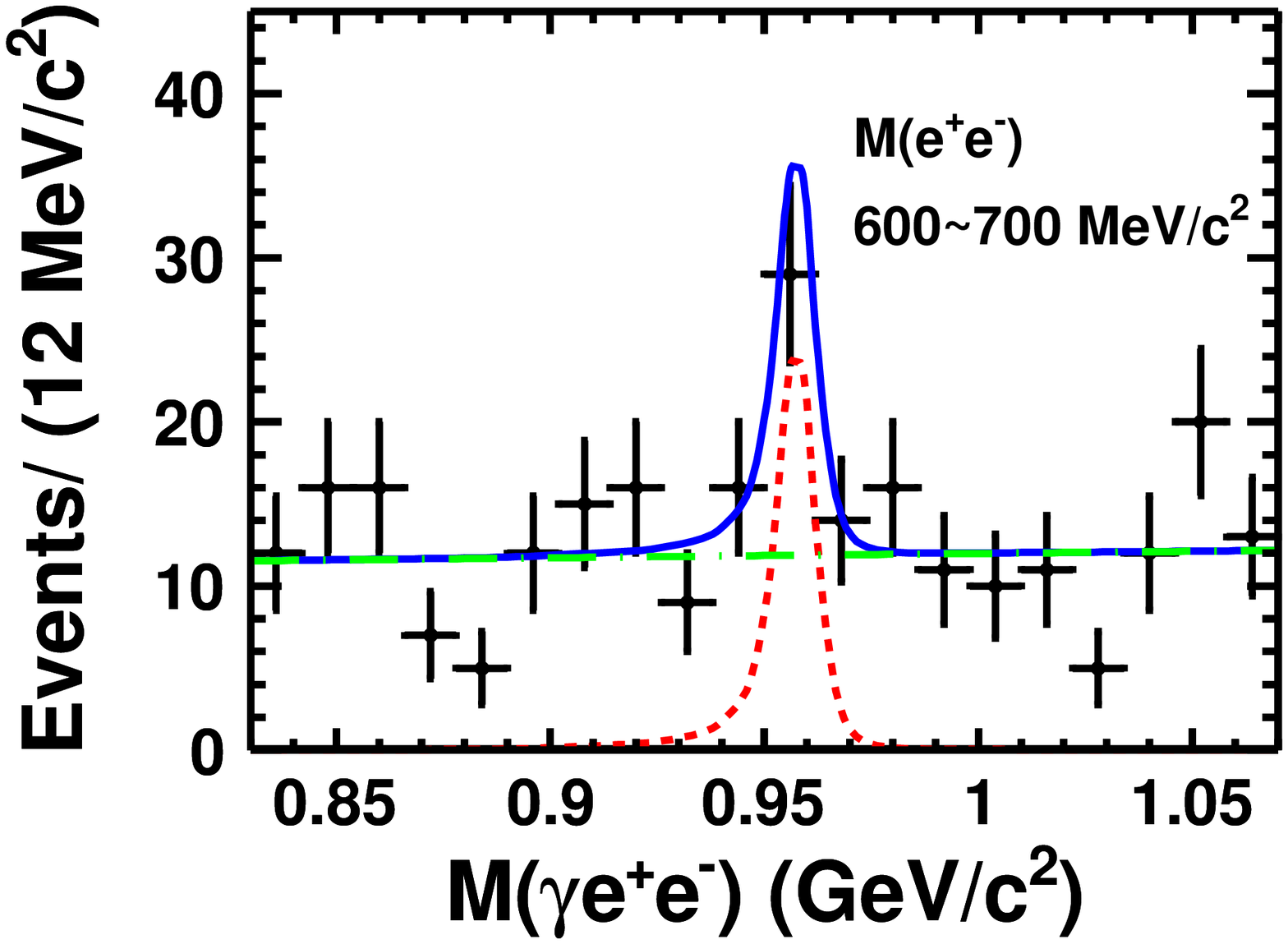}}\put(-120,65)
  \subfigure{
    \includegraphics[width=0.3\linewidth]{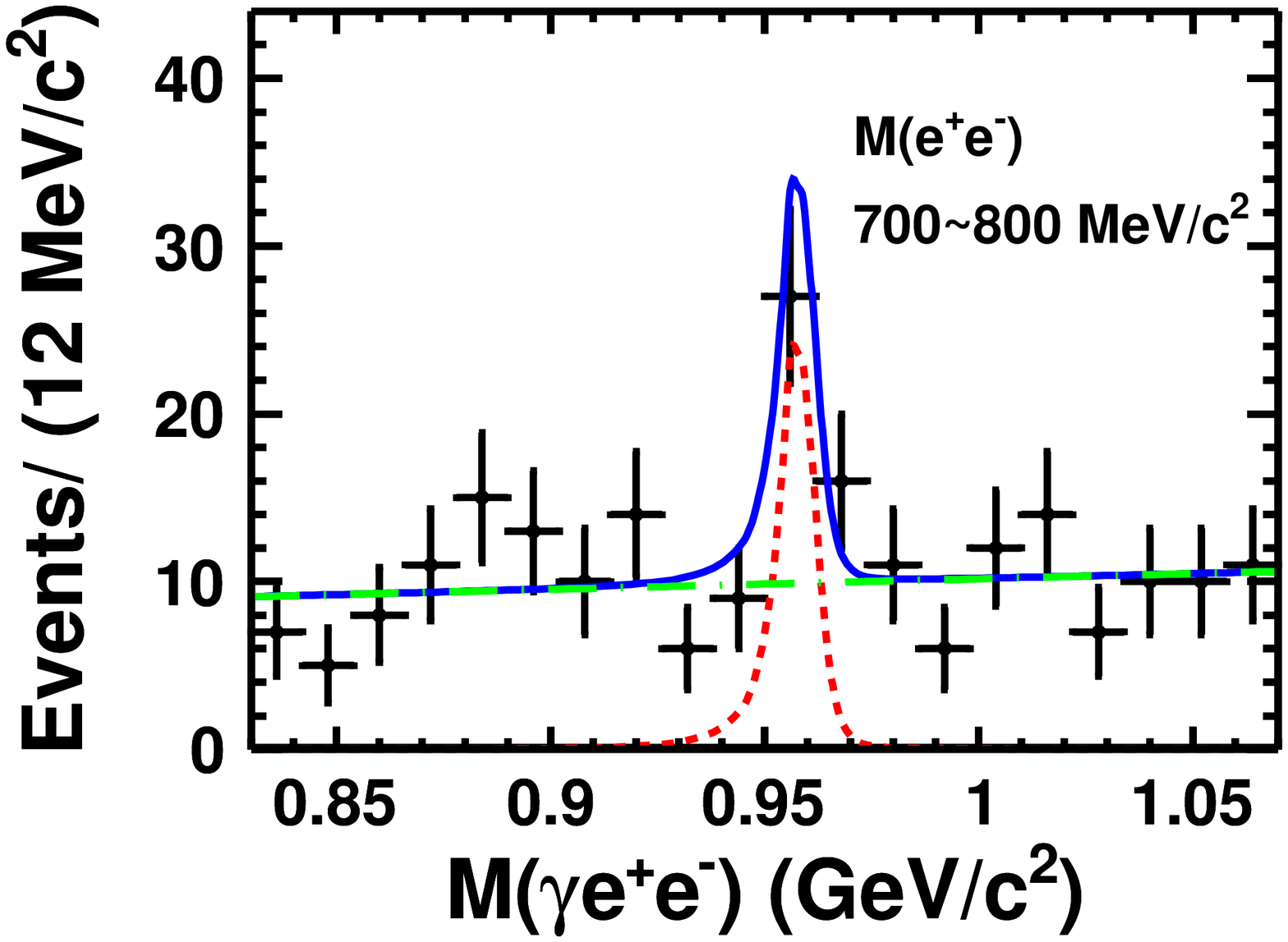}}\put(-120,65){}
  \caption{Results from bin-by-bin fits to the $M(\gamma e^+ e^-)$ distributions for different $M(e^+e^-)$ bins.
            The (black) crosses are data, the (red) dashed curves represent the signal, the (green) dot-dashed
            curves show the non-peaking backgrounds, the (orange) shaded component for the $M(e^+e^-)<100$~\MeV
            bin is the shape of the peaking background from $J/\psi\to\gamma\eta', \eta'\to\gamma\gamma$.
            The total fit results are shown as (blue) solid curves.}
  \label{binFit}
\end{figure*}

\begin{figure}[hbtp]
\centering
    \includegraphics[width=0.9\linewidth]{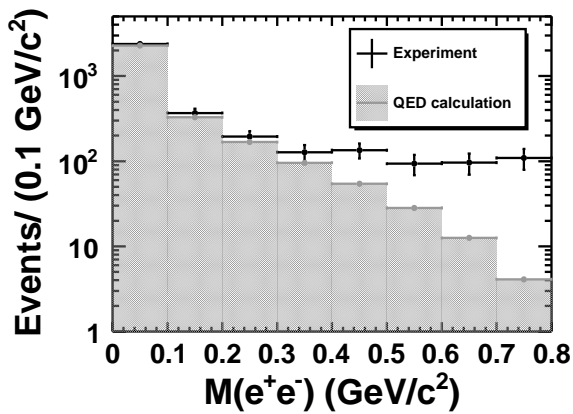}
    \caption{ Efficiency-corrected signal yields $n^{\text{corr}}_i$ versus $M(e^+e^-)$. The (black)
              crosses are data and the (gray) shaded histogram indicates the point-like QED result.}
    \label{EffCorrected}
\end{figure}

\begin{figure}[hbtp]
\begin{minipage}[b]{\linewidth}
\centering
\includegraphics[width=0.9\linewidth]{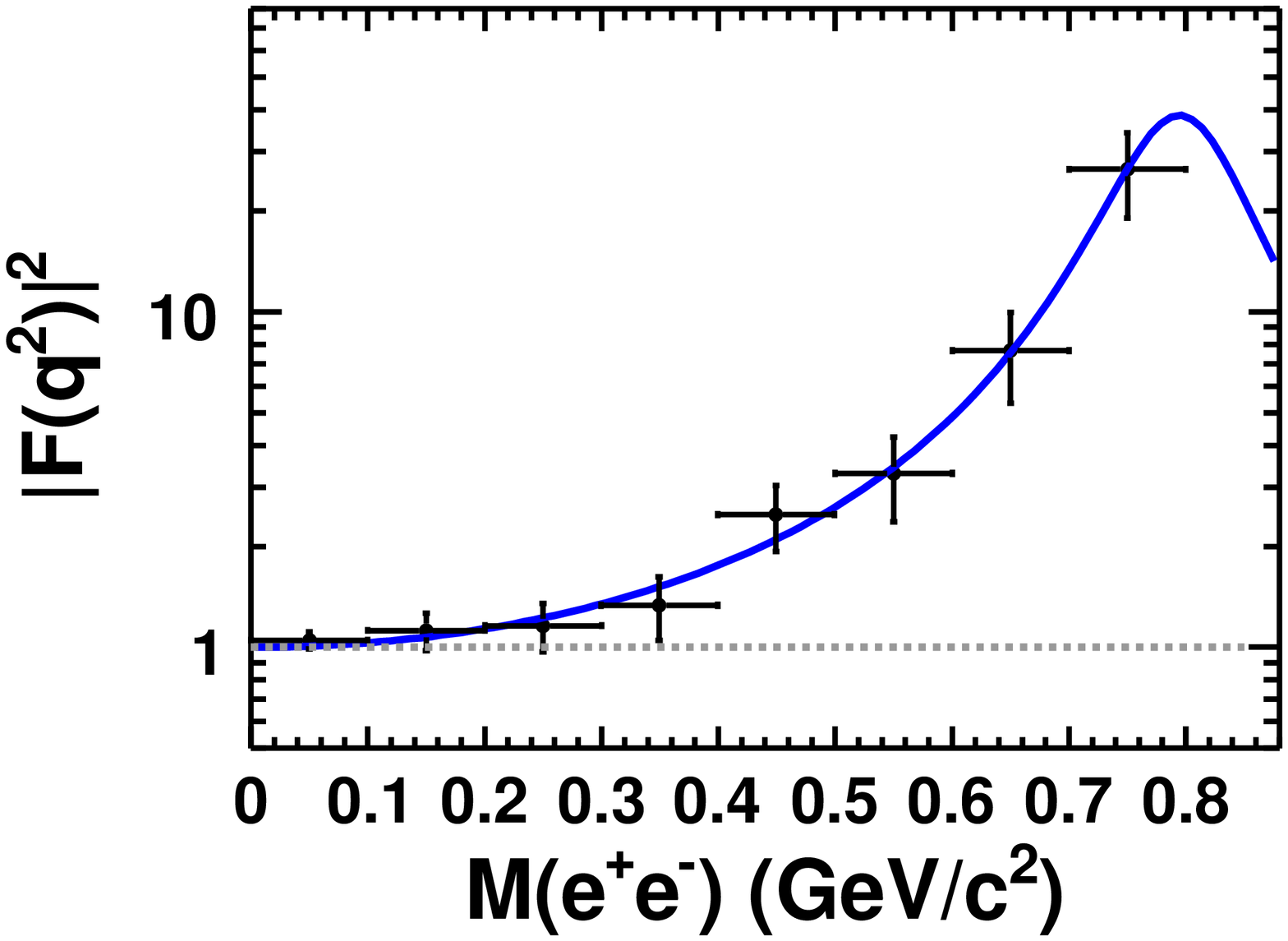}
    \caption{Fit to the single pole form factor  $|F|^2$ using Eq.~(\ref{FF2}). The (black) crosses
              are data, where the statistical and systematic uncertainties are
              combined, the (blue)
              solid curve shows the fit results. The (gray) dotted line shows to the point-like
              case (\emph{i.e.} with $|F|^2=1$) for comparison. }
    \label{etap:FF1total}
\end{minipage}
\hfill
\begin{minipage}[b]{\linewidth}
\centering
    \includegraphics[width=0.9\linewidth]{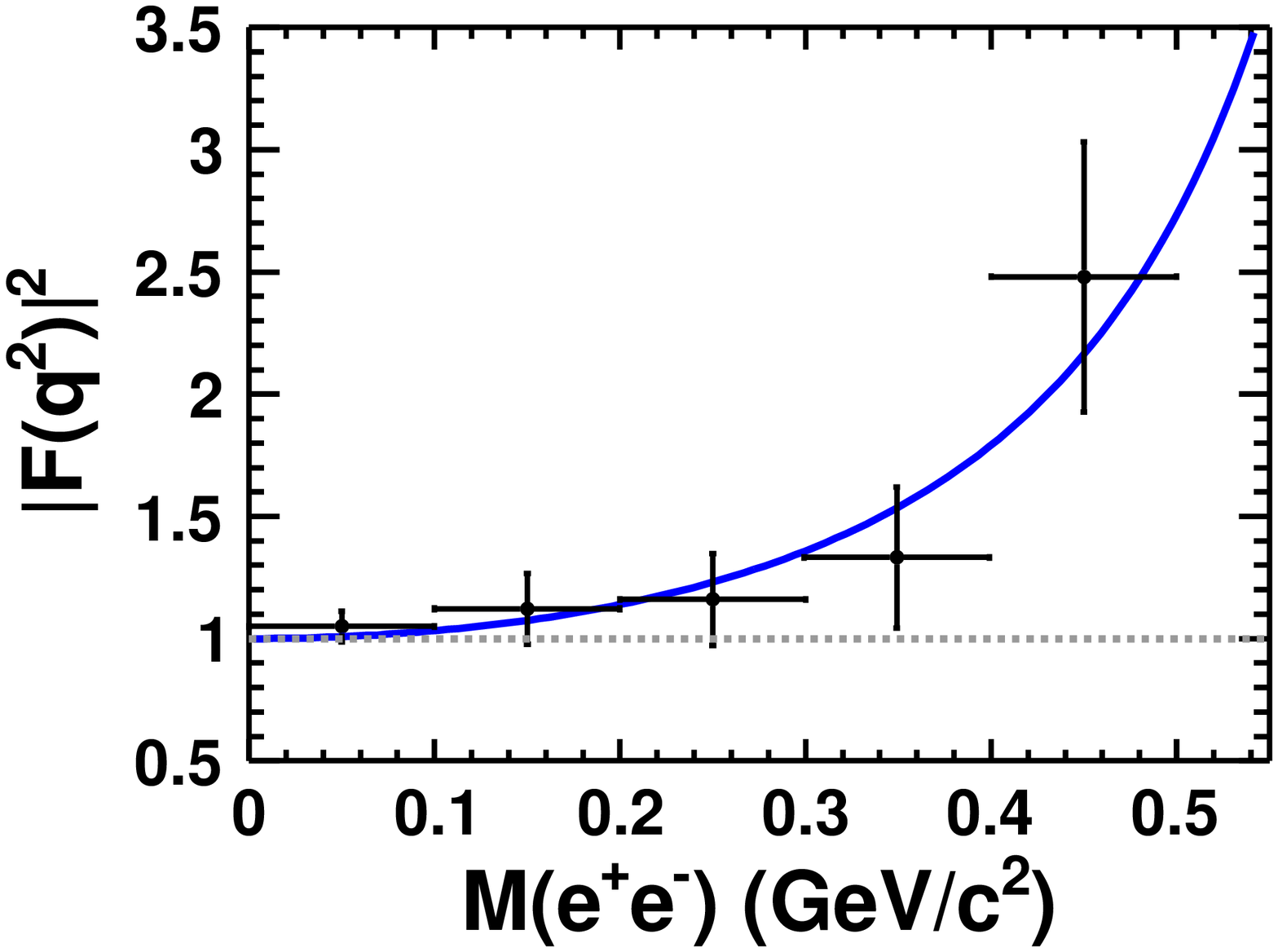}
    \caption{Determination of the form factor slope by fitting to $|F|^2$ using the single pole form factor of
             Eq.~(\ref{simplepole}). The (black) crosses are data, where the uncertainties are
             the combined statistical and systematic uncertainties, the (blue) solid curve shows the fit result. The
             (gray) dotted line corresponds to the point-like particle case (with $|F|^2=1$).}
    \label{etap:FF2total}
\end{minipage}
\end{figure}

\section{Form factor measurement} \label{FFM}

The TFF is extracted from the bin-by-bin efficiency corrected signal yields for eight different $M(e^+e^-)$ bins.
The bin widths are all chosen to be 0.1~\GeV.  Since this is much wider than the $M(e^+e^-)$ resolution,
which is 5$\sim$6~\MeV~depending on $M(e^+e^-)$, no unfolding is needed.
 The signal yield in each $M(e^+e^-)$ bin $i$ is obtained by performing bin-by-bin fits to the
$M(\gamma e^+e^-)$ mass distributions using the fitting procedure
described in Section~\ref{gammaee}. The peaking background from the
$J/\psi\to\gamma\eta', \eta'\to\gamma\gamma$ only exists in the first
bin, and the yield is fixed to the normalized number in the bin.
The fit results are shown in Fig.~\ref{binFit}.  The fitted ($n^{\text{obs}}_i$) and efficiency-corrected
signal yields ($n^{\text{corr}}_i$) for each $M_i(e^+e^-)$ bin are summarized in Table~\ref{binyields}.
Figure~\ref{EffCorrected} shows the efficiency-corrected signal yields versus $M(e^+e^-)$ with the QED
shape superimposed for comparison. The discrepancy between QED and data, which reflects the TFF, is
evident in the high $M(e^+e^-)$ region.

The systematic uncertainties on $n^{\text{corr}}_i$ include the uncertainty
from the MDC tracking efficiency, PID, photon detection,
kinematic fit, veto of gamma conversion, background description and
signal shape; they are the same
as those described in Section~\ref{systemtic:section}.

The partial ratio $r_i = \Delta \Gamma(\eta^\prime \to \gamma e^+e^-)_i /\Gamma(\eta^\prime \to \gamma \gamma)$
 for each given $M(e^+e^-)$ bin $i$,  is defined as
 \begin{eqnarray}
  \label{partial-ratio}
 r_i \equiv  \frac{\Delta \Gamma(\etap \to \gamma e^+ e^-)_i}{\Gamma(\etap\to\gamma\gamma)}&= &\frac{n^{\text{corr}}_i  \epsilon_{\etap\to\gamma\gamma}} {N_{\etap\to\gamma\gamma}} ,
 \end{eqnarray}
where $\Delta\Gamma(\eta^\prime \to \gamma e^+e^-)_i $ is the integrated rate in each  $M(e^+e^-)$ interval.

The result for $|F|^2$ in each $M(e^+e^-)$ bin is obtained by dividing
the value $r_i$ by the integrated QED predication in each $M(e^+e^-)$ interval (see Eq.~(\ref{eq:decaywidth})).
The values of $|F|^2$ for each $M(e^+e^-)$ bin are summarized in Table~\ref{fq2bin}.

\begin{table*}[hbtp]
  \caption{Fitted ($n^{\text{obs}}_i$) and efficiency-corrected
    ($n^{\text{corr}}_i$) signal yields
    for the eight $M(e^+e^-)$ bins, and ratios ($r_i$). The uncertainties are  statistical only. }
  \newcommand{\tabincell}[2]{\begin{tabular}{@{}#1@{}}#2\end{tabular}}
  \label{binyields}
  \begin{center}
     \renewcommand{\arraystretch}{2}
     \begin{tabular}{ccccc}
        \hline\hline
         $M(e^+e^-)$~(\GeV) \hspace{0.3cm} & [0.0,  0.1]   & [0.1, 0.2] & [0.2, 0.3] & [0.3,  0.4]   \\ \hline

           \hspace{0.3cm}  $n^{\text{obs}}_i$  \hspace{0.3cm} & \hspace{0.3cm}  $545\pm27$  \hspace{0.3cm}  &$86.5\pm10.7$  \hspace{0.3cm} & $62.1\pm9.8$  \hspace{0.3cm}  &$45.6\pm9.7$  \hspace{0.3cm}    \\
     \hspace{0.3cm}  $n^{\text{corr}}_i$  \hspace{0.3cm}            & \hspace{0.3cm}  $2380\pm120$  \hspace{0.3cm}  &$368\pm46$  \hspace{0.3cm} & $194\pm31$  \hspace{0.3cm}  &$128\pm27$  \hspace{0.3cm}    \\
      \hspace{0.3cm}  $r_i~(10^{-2})$ \hspace{0.3cm}            & \hspace{0.3cm}  $1.44\pm0.07$  \hspace{0.3cm}  &$0.22\pm0.03$  \hspace{0.3cm} & $0.12\pm0.02$  \hspace{0.3cm}  &$0.08\pm0.02$  \hspace{0.3cm}    \\ \hline

         $M(e^+e^-)$~(\GeV) \hspace{0.3cm}  & [0.4, 0.5]  & [0.5, 0.6] & [0.6, 0.7]  & [0.7, 0.8] \\ \hline
           \hspace{0.3cm}  $n^{\text{obs}}_i$  \hspace{0.3cm}            & \hspace{0.3cm}  $45.4\pm9.0$  \hspace{0.3cm}  &$29.9\pm8.0$  \hspace{0.3cm} & $28.0\pm7.8$  \hspace{0.3cm}  &$25.0\pm6.9$  \hspace{0.3cm}    \\
     \hspace{0.3cm}  $n^{\text{corr}}_i$  \hspace{0.3cm}            & \hspace{0.3cm}  $135\pm27$  \hspace{0.3cm}  &$93.3\pm25.0$  \hspace{0.3cm} & $96.2\pm26.8$  \hspace{0.3cm}  &$109\pm30$  \hspace{0.3cm}    \\
      \hspace{0.3cm}  $r_i~(10^{-2})$ \hspace{0.3cm}            & \hspace{0.3cm}  $0.08\pm0.02$  \hspace{0.3cm}  &$0.06\pm0.02$  \hspace{0.3cm} & $0.06\pm0.02$  \hspace{0.3cm}  &$0.07\pm0.02$  \hspace{0.3cm}    \\ \hline
        \hline
      \end{tabular}
  \end{center}
\end{table*}

A variety of models have been traditionally used to parameterize the TFF. The most common one,
based on VMD~\cite{JJS}, uses only the first term in the dispersion relation. In this single
pole model, the TFF is given by Eq.~(\ref{FF2}). The results of a least-squares fit
with the single pole model is shown in Fig.~\ref{etap:FF1total}; the parameters of the
form factors are determined to be
$\Lambda_{\eta'} = (0.79\pm0.05)$~GeV, $\gamma_{\eta'} = (0.13\pm0.06)$~GeV.
From the fitted value of the parameter $\Lambda_{\eta'}$, the slope of the form factor is obtained
to be $(1.60\pm0.19)$~GeV$^{-2}$, in agreement with the result
$b_{\eta'} = (1.7\pm0.4)$ GeV$^{-2}$ obtained in the process of
$\eta' \to \gamma \mu^+\mu^-$~\cite{Landsberg}.

\begin{table*}[hbtp]
  \caption{Values of $|F|^2$ in each $M(e^+e^-)$ bin, where the first uncertainties are statistical and
            the second ones systematic.}
  \newcommand{\tabincell}[2]{\begin{tabular}{@{}#1@{}}#2\end{tabular}}
  \label{fq2bin}\footnotesize
  \begin{center}
     \renewcommand{\arraystretch}{2}
     \begin{tabular}{c|cccc}
        \hline\hline
         $M(e^+e^-)$~(\GeV) \hspace{0.3cm} & [0.0,  0.1]   & [0.1, 0.2] & [0.2, 0.3] & [0.3,  0.4]   \\ \hline

           \hspace{0.3cm} $|F|^2$   \hspace{0.3cm}            & \hspace{0.3cm}  $1.05\pm0.05\pm0.03$  \hspace{0.3cm}  &$1.12\pm0.14\pm0.04$  \hspace{0.3cm} & $1.16\pm0.18\pm0.05$  \hspace{0.3cm}  &$1.33\pm0.28\pm0.05$  \hspace{0.3cm}    \\ \hline
         $M(e^+e^-)$~(\GeV) \hspace{0.3cm}  & [0.4, 0.5]  & [0.5, 0.6] & [0.6, 0.7]  & [0.7, 0.8] \\ \hline
           \hspace{0.3cm} $|F|^2$  \hspace{0.3cm}    &  \hspace{0.3cm} $2.48\pm0.49\pm0.25$ \hspace{0.3cm}  & $3.30\pm0.88\pm0.31$ \hspace{0.3cm}  & $7.66\pm2.13\pm0.89$ \hspace{0.3cm}  & $26.6\pm7.3 \pm1.9$    \hspace{0.3cm}    \\
        \hline\hline
      \end{tabular}
  \end{center}
\end{table*}

To test the robustness of the slope extracted from the simple pole model, we also fit the
data below 0.5~GeV/$c^2$ using the single pole Ansatz used in lighter meson studies:
\begin{eqnarray}\label{simplepole}
   F(q^2)=\frac{1}{(1-q^2/\Lambda^2)},
\end{eqnarray}
The parameterization diverges at $M(e^+e^-)=\Lambda$ and, therefore, can not be used for the
whole kinematic region. The result of this fit is shown in Fig.~\ref{etap:FF2total}. The slope of
the form factor is determined to be $b_{\eta'} = (1.58\pm0.34)$~GeV$^{-2}$, which is in good
agreement with the  result of $(1.60\pm0.19)$~GeV$^{-2}$
using Eq.~(\ref{FF2}).

The quadratic difference between the uncertainties of the parameters
 with only statistical errors used in the fits and the
uncertainties of the parameters with combined statistical
and systematic errors used in the fits is taken as the systematic uncertainty on the parameters.
The resulting parameters
in Eq.~(\ref{FF2}) are determined to be $\Lambda_{\eta'} = (0.79\pm0.04(\text{stat.})\pm0.02(\text{sys.}))$~GeV,
$\gamma_{\eta'} = (0.13\pm0.06(\text{stat.})\pm0.03(\text{sys.}))$~GeV, respectively.

\section{Summary}

In summary, with a sample of 1.31 billion $J/\psi$ events collected in the BESIII detector, we
have made the first measurement of the EM Dalitz decay process $\etap \to \gamma e^+e^-$ and measure
the ratio $\Gamma(\etap \to \gamma e^+e^-)/\Gamma(\etap\to\gamma\gamma)=
(2.13\pm0.09(\text{stat.})\pm0.07(\text{sys.}))\times10^{-2}$.  Using the PDG value for the
 $\eta'\to \gamma \gamma$ branching fraction~\cite{PDG}, we determine
${\cal B}(\eta' \to \gamma e^+e^-)= (4.69 \pm0.20(\text{stat.})\pm0.23(\text{sys.}))\times10^{-4}$.
We present measurements of the TFF as a function of $M(e^+e^-)$.
Our TFF results can be described with a single pole parameterization Eq.~(\ref{FF2}), with
mass and width parameters of $\Lambda_{\eta'} = (0.79\pm0.04(\text{stat.})\pm0.02(\text{sys.}))$~GeV,
and $\gamma_{\eta'} = (0.13\pm0.06(\text{stat.})\pm0.03(\text{sys.}))$~GeV, respectively.
The slope of the TFF corresponds to
$(1.60\pm0.17(\text{stat.})\pm0.08(\text{sys.}))$~GeV$^{-2}$ and agrees within errors with the VMD model
predictions.
The uncertainty of the $\eta'$ transition form factor slope matches the best
determination in the space-like region from the CELLO collaboration
$b_{\eta'} = (1.60\pm0.16)$~GeV$^{-2}$~\cite{CELLO},
and improves the previous determination of the slope in the time-like
region $b_{\eta'} = (1.7\pm0.4)$~GeV$^{-2}$~\cite{Landsberg, etapGmu}.
The $\eta'$ form factor is determined by both
universal $\pi^+\pi^-$ rescattering and a reaction specific part, with the latter
contributing about 20\% to the form factor slope~\cite{M1309}. Therefore our
result is sensitive specifically to the $\eta'$ internal EM structure.
In addition, the decay $\eta'\to \gamma e^+e^-$ is closely related to $\eta'\to\gamma\pi^+\pi^-$, and in
particular the transition form factor could be predicted from the invariant mass
distribution of the two pions and the branching ratio of the
$\eta'\to\gamma\pi^+\pi^-$ decay in a model independent way using a dispersive integral.
Also, the knowledge of the TFF is useful for studies of the HLbL
scattering contribution to the muon anomalous magnetic moment, $a_\mu = (g_\mu -2)/2$~\cite{ Blum:2013xva}.

\section{ACKNOWLEDGMENT}
The BESIII collaboration thanks the staff of BEPCII and the IHEP computing center for their strong support. This work is supported in part by National Key Basic Research Program of China under Contract No. 2015CB856700; National Natural Science Foundation of China (NSFC) under Contracts Nos. 11125525, 11235011, 11322544, 11335008, 11425524; the Chinese Academy of Sciences (CAS) Large-Scale Scientific Facility Program; Joint Large-Scale Scientific Facility Funds of the NSFC and CAS under Contracts Nos. 11179007, U1232201, U1332201; CAS under Contracts Nos. KJCX2-YW-N29, KJCX2-YW-N45; 100 Talents Program of CAS; INPAC and Shanghai Key Laboratory for Particle Physics and Cosmology; German Research Foundation DFG under Contract No. Collaborative Research Center CRC-1044; Istituto Nazionale di Fisica Nucleare, Italy; Ministry of Development of Turkey under Contract No. DPT2006K-120470; Russian Foundation for Basic Research under Contract No. 14-07-91152; U. S. Department of Energy under Contracts Nos. DE-FG02-04ER41291, DE-FG02-05ER41374, DE-FG02-94ER40823, DESC0010118; U.S. National Science Foundation; University of Groningen (RuG) and the Helmholtzzentrum fuer Schwerionenforschung GmbH (GSI), Darmstadt; WCU Program of National Research Foundation of Korea under Contract No. R32-2008-000-10155-0


\end{document}